\documentclass[letter,traditabstract]{aa} 

\usepackage{graphicx}
\usepackage{txfonts}
\begin{document}

\title{Light curves of the latest FUor: Indication of a close binary
}

\author{
  M. Hackstein
  \inst{1}
  \and
  M. Haas
  \inst{1}
  \and
  \'A. K\'osp\'al
  \inst{2}		
  \and
  F.-J. Hambsch
  \inst{3}
  \and
  R. Chini
  \inst{1,4}
  \and
  P. \'Abrah\'am
  \inst{2}
  \and
  A. Mo\'or
  \inst{2}
  \and
  F. Pozo Nu\~nez
  \inst{1}
  \and \\
  M. Ramolla
  \inst{1}
  \and
  Ch. Westhues
  \inst{1}
  \and
  L. Kaderhandt
  \inst{1}
  \and
  Ch. Fein
  \inst{1}
  \and
  A. Barr Dom\'inguez
  \inst{4}
  \and
  K.-W. Hodapp
  \inst{5}
}
\institute{
  Astronomisches Institut, Ruhr--Universit\"at Bochum,
  Universit\"atsstra{\ss}e 150, 44801 Bochum, Germany
  \and
  Konkoly Observatory, Research Centre for Astronomy and Earth
  Sciences, Hungarian Academy of Sciences, P.O. Box 67, H-1525
  Budapest, Hungary
  \and
  Center for Backyard Astrophysics (Antwerp), American Association of
  Variable Star Observers (AAVSO), Verenigin Voor Sterrenkunde (VVS),
  Andromeda Observatory, Oude Bleken 12, 2400 Mol, Belgium
  \and
  Instituto de Astronomia, Universidad Cat\'{o}lica del
  Norte, Avenida Angamos 0610, Casilla
  1280 Antofagasta, Chile
  \and
  Institute for Astronomy, University of Hawaii,
  640 N. Aohoku Place, Hilo HI 96720, USA
}

\date{Received ; accepted}


\abstract{
  We monitored the recent FUor 2MASS J06593158-0405277 (V960\,Mon) since November
  2009 at various observatories and multiple wavelengths. After the
  outburst by nearly 2.9\,mag in $r$ around September 2014 the brightness
  gently fades until April 2015 by nearly 1\,mag in $U$ and 0.5\,mag in
  $z$. Thereafter the brightness at $\lambda>5000\,\AA$~ was constant until
  June 2015 while the shortest wavelengths ($U, B$) indicate a new rise,
  similar to that seen for the FUor V2493 Cyg (HBC722).
  Our near-infrared (NIR) monitoring between December 2014 and April 2015
  shows a smaller outburst amplitude ($\sim$2\,mag) and a smaller
  (0.2\,--\,0.3\,mag) post-outburst brightness decline.
  Optical and NIR color-magnitude diagrams indicate that
  the brightness decline is caused by growing extinction.
  The post-outburst light curves are modulated by an oscillating
  color-neutral pattern with a period of about 17 days and an
  amplitude declining from $\sim$0.08\,mag in October 2014 to
  $\sim$0.04\,mag in May 2015.
  The properties of the oscillating pattern lead us to suggest the
  presence of a close binary with eccentric orbit.
}

\maketitle
%
\section{Introduction}

Recently \cite{Maehara14} detected a FUor-type brightness
outburst in 2MASS J06593158-0405277, henceforth denoted V960\,Mon \citep{Semkov15}.

\cite{Hillenbrand14} confirmed the optical spectroscopic
results on the P-Cygni line profiles and determined the
temperature and gravity.
Subsequent near-infrared (NIR) spectra by \cite{Reipurth15}
and \cite{Pyo15} corroborated the FUor nature.
These spectra and X-ray spectra by \cite{Pooley15}
indicated low extinction towards the FUor.
\cite{Hackstein14} presented the first multiband monitoring results,
revealing that the outburst was about 1\,mag brighter than reported by
\cite{Maehara14}. \cite{Varricatt15} reported on the NIR
monitoring during December 2014.

After the flood of initial Astronomical Telegrams, \cite{Kospal15}
presented an almost complete picture of the FUor progenitor.
It is likely a young Class II source
associated with a group of several other YSOs.
The FUor progenitor star has a temperature of $T_{eff} = 4000$\,K, a mass of
0.75\,$M_\odot$ and an age of about 6$\times$10$^{5}$\,yrs.

\cite{Caratti15} reported on NIR adaptive optics imaging and
spectroscopy using VLT/SINFONI. An extended 90\,AU disk-like structure
was resolved, as were two companions with projected distances of 11
and 100\,AU from V960\,Mon. Both companions display accretion
signatures.

Here we present the first results from multifilter monitoring of V960\,Mon.
We focus on the full outburst amplitude, the nature
of the subsequent brightness decline, and the
superimposed color-neutral brightness oscillations.

\section{Observations and data}

{\it Optical monitoring} was carried out at three observatories:

{\it Universit\"atssternwarte Bochum (USB),
Chile:}
Between November 2009 and May 2015 we used
the robotic 15\,cm Twin Telescope RoBoTT near Cerro
Armazones\footnote{http://www.astro.ruhr-uni-bochum.de/astro/oca/}.
The observations were obtained in the course of the Bochum Galactic
Disk Survey, described in detail by
\cite{Haas12} and  \cite{Hackstein15}.
V960\,Mon was monitored in the Sloan {\it r} and {\it i} filters, and since
October 2014 additionally in the Johnson {\it UBV} and Sloan {\it z} filters
with median sampling of 1~day.
The light curves are created
using several hundred non-variable stars
located in the same field and of similar brightness
to V960\,Mon.
The absolute photometric calibration is based on several
standard star fields measured each night.

{\it Konkoly Observatory, Hungary: }
Between December 2014 and March 2015, we monitored V960\,Mon in the
Johnson-Cousins $BVRI$ filters using the 60/90/180\,cm (aperture diameter/primary mirror diameter/focal length)
Schmidt telescope. Details about the instrument and the steps of data reduction and photometry
are described in \cite{Kospal11}. We used 110 stars within 10$'$ of the
source as comparison stars. Photometric calibration was done using the
UCAC4 catalog magnitudes of the comparison stars \citep{Zacharias13},
where we first converted the Sloan $r$ and $i$ magnitudes to
Johnson-Cousins $R$ and $I$ magnitudes using the formulae of \cite{Jordi06}.

{\it The Remote Observatory Atacama Desert (ROAD): }
We performed extensive monitoring
using the robotic 40~cm telescope
near San Pedro de Atacama \citep{Hambsch12}.
V960\,Mon was observed in the Johnson-Cousins $BVI$ filters
between December 2014 and
June 2015 twice per night with median
sampling of 1 day.
Photometry was obtained relative to two non-variable
AAVSO stars
located in the same field and which have a similar brightness
to V960\,Mon.

{\it Near-IR monitoring} was performed in $J,H$, and $K_{s}$
between December 2014 and April 2015 using the 0.8\,m
Infrared Imaging System \citep[IRIS,][]{Hodapp10} at
USB. Images were obtained and reduced in the standard manner.
Photometry was obtained relative to 20
non-variable high-quality flag (AAA) 2MASS stars
located in the same field.

All light curves are listed in the online table
with the following columns: telescope, date, filter, mag, err.
For Sloan filters we use the AB system, for all other filters the Vega
mag system. Photometric errors are
smaller than 0.015\,mag for RoBoTT and ROAD, about 0.02\,mag for IRIS,
about 0.03\,--\,0.04\,mag for Konkoly,
and about 0.05\,--\,0.1\,mag  for the RoBoTT $U$-band data.

\begin{figure}
  \hspace{-5mm}
   \includegraphics[width=1.1\columnwidth, angle=0]{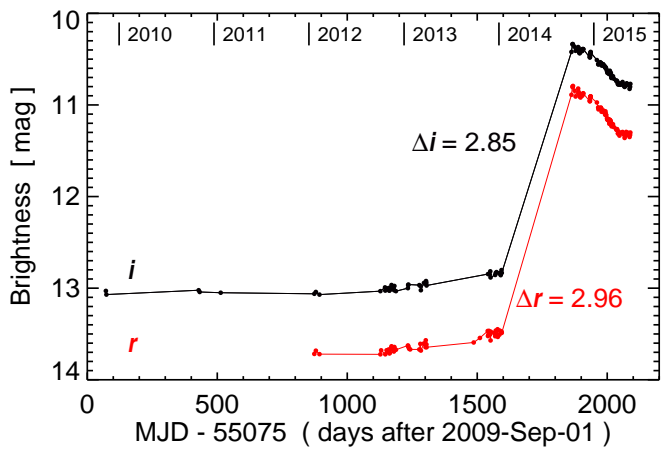}
  \caption{Sloan $r$ and $i$ light curves of V960\,Mon
    obtained at USB Bochum (USB).
  }
  \label{fig_v6_lc}
\end{figure}

\begin{figure}
  \vspace{-1mm}
  \hspace{-5mm}
  \includegraphics[width=1.1\columnwidth, angle=0]{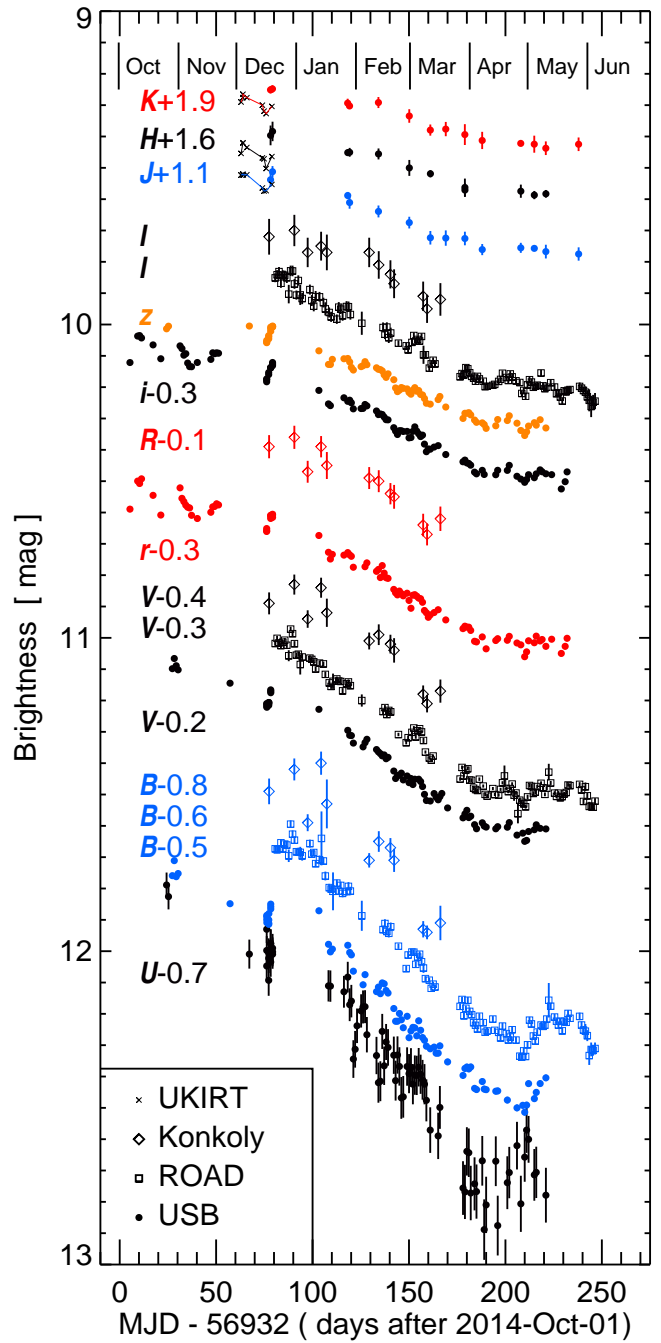}
  \caption{Post-outburst light curves of V960\,Mon.
  }
  \label{fig_all_lc}
\end{figure}

\section{Results and discussion}

\subsection{Optical and infrared light curves}

Figure \ref{fig_v6_lc} shows the $r$ and $i$ light curves.
Between 11 November 2009 and 17 April 2012, the sparse light curves
did not catch any variations larger than 0.05\,mag.
Thereafter, the denser sampled light curves (median $<$2 days) show a
scatter of up to 0.2\,mag amplitude on the time scale of days and a
gradual brightness increase until 11 January 2014 to $i$ = 12.77\,mag and $r$ =
13.42\,mag.
When the target became visible again, on 6 October 2014 the brightness had
drastically increased by about 2.6\,mag to $i$ =10.2\,mag and $r$ = 10.75\,mag.
The color before the
outburst $r-i$ = 0.65 $\pm$ 0.04\,mag became slightly bluer during and after
the outburst $r-i$ = 0.55 $\pm$ 0.03\,mag.

The observations confirm the outburst of V960\,Mon,
where a brightening of $I_{c}$ = 1.5\,mag was reported in
the discovery telegram by \cite{Maehara14}.
In contrast to that relatively modest brightening, we find a much
stronger brightness outburst by about 2.6\,mag between 11 January and
6 October 2014, and in total about 2.9\,mag between
2009\,--\,2011 and October 2014.
This is still below the typical 5\,mag brightening of known FUors.

\begin{figure*}
  \vspace{-1mm}
  \hspace{-6mm}
  \includegraphics[width=6.7cm, angle=0]{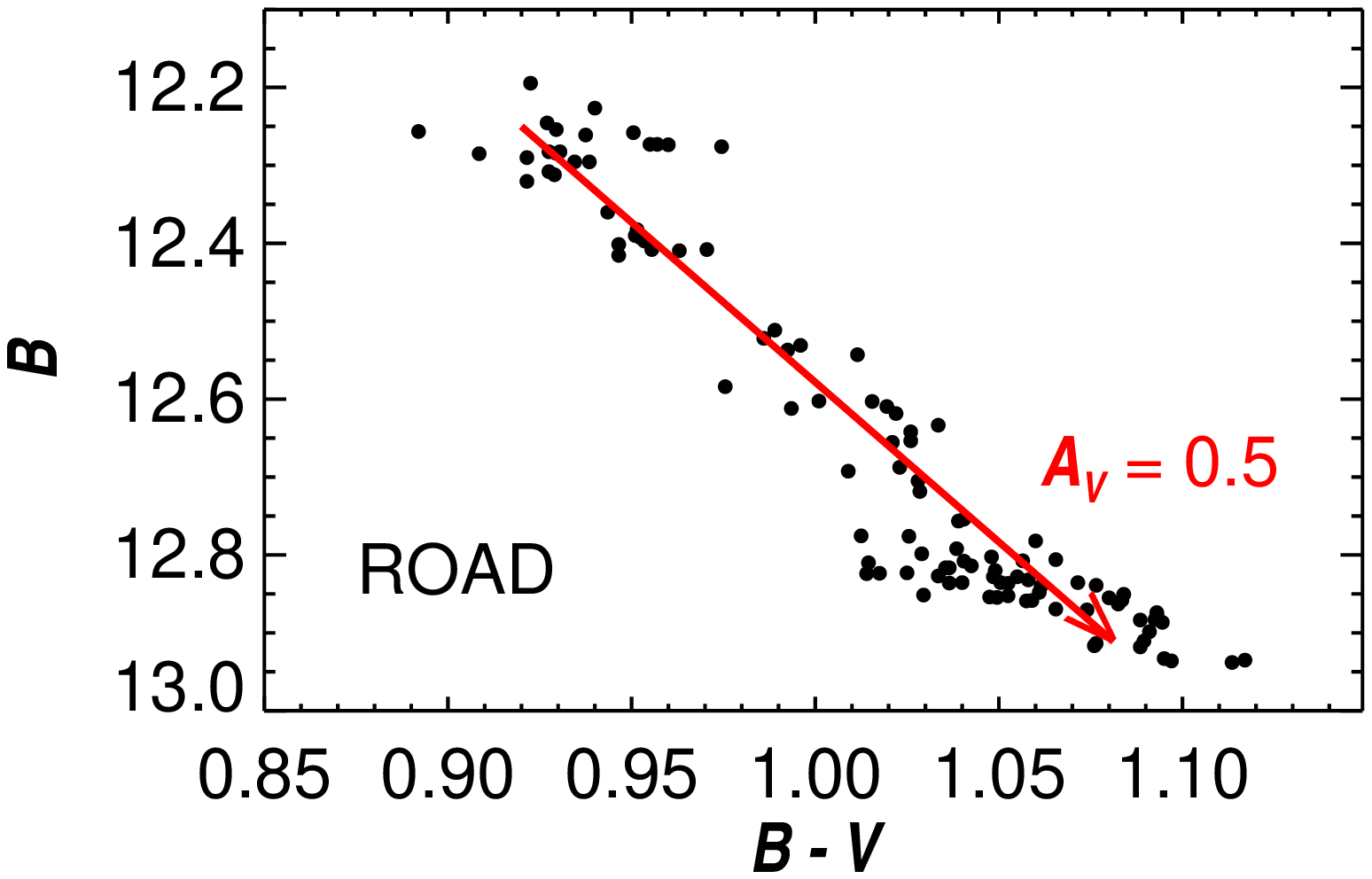}
  \hspace{-6mm}
  \includegraphics[width=6.7cm, angle=0]{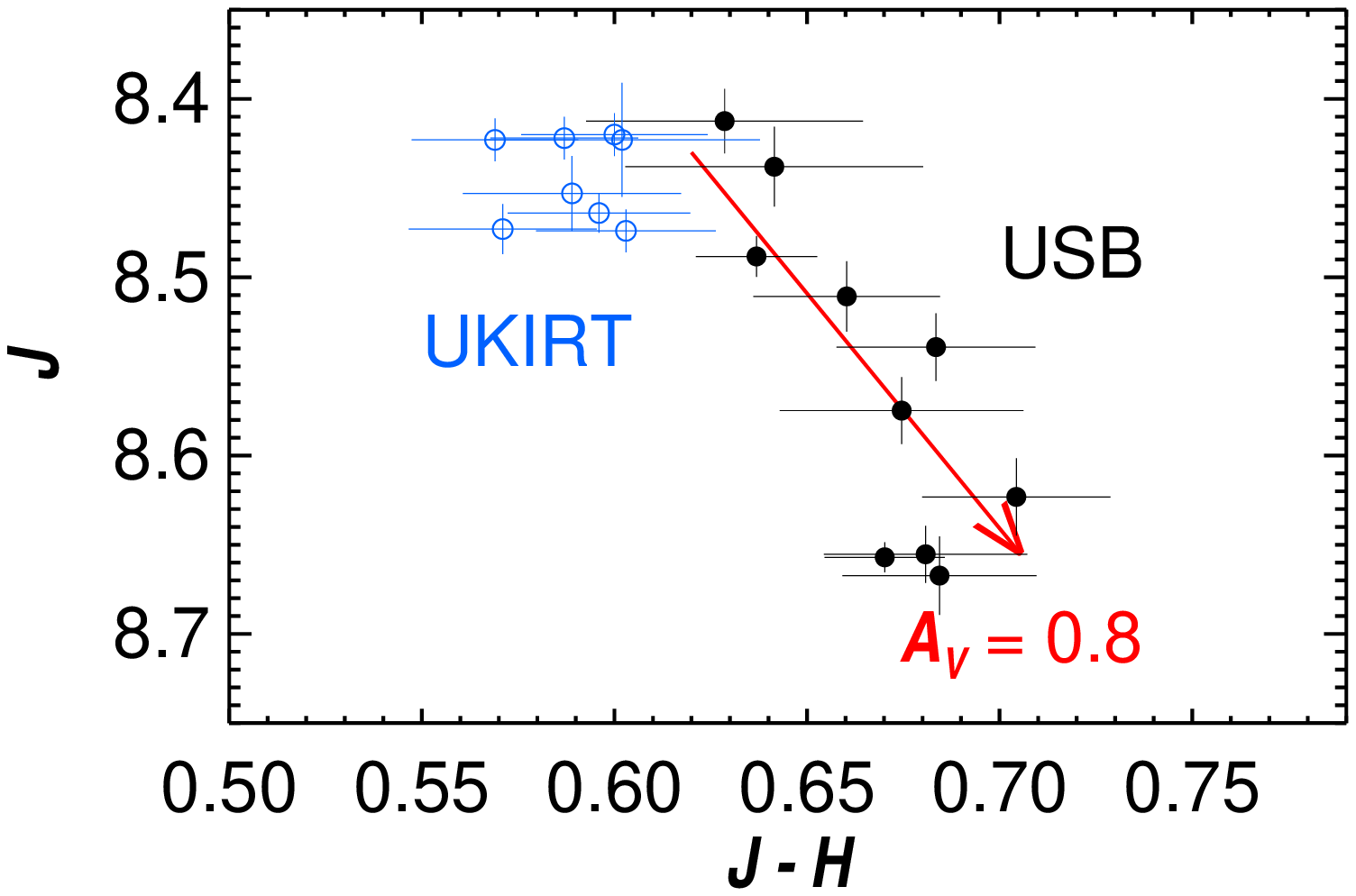}
  \hspace{-6mm}
  \includegraphics[width=6.7cm, angle=0]{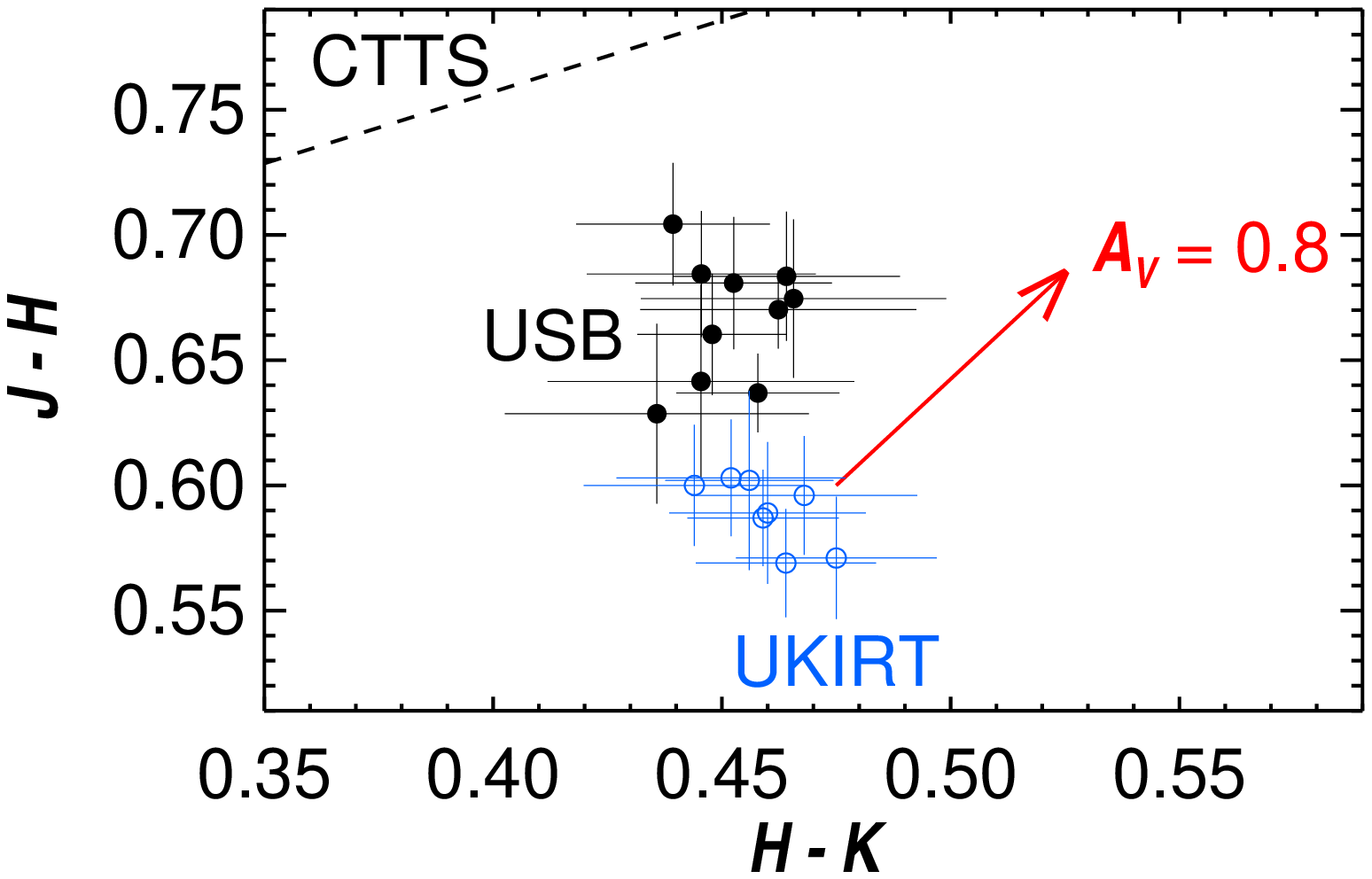}
  \caption{
    Post-outburst color-magnitude diagrams at
    optical and NIR wavelengths.
    The UKIRT data are from \cite{Varricatt15}.
    The red arrow shows
    $A_{V}$ as labeled (using
    standard interstellar reddening, \citealt{rieke85}), the dashed line indicates the domain of classical T\,Tauri stars.
  }
  \label{fig_cmds}
\end{figure*}

Figure \ref{fig_all_lc}
shows the light curves in various bands after the outburst.
The main features are as follows:

Between October 2014 and April 2015 the brightness
faded by nearly 1\,mag in $U$ and 0.5\,mag in $z$.
Thereafter the brightness
was constant at wavelengths longer than $V$,
while $U$ and $B$ show a turn-over between April and May 2015; this
indicates the transition to a brightness rise
similar to that seen for V2493~Cyg (\citealt{Semkov14}).
Until 4 September 2015, however, when V960\,Mon became visible again,
no strong rise in $BVRI$ has been found
($B$=13.14, $V$=11.87, $R$=11.04, $I$=10.14, \citealt{Semkov15}).

Our NIR monitoring between December 2014 and April 2015
shows a smaller outburst amplitude (about 2\,mag compared to 2MASS, see \citealt{Kospal15}) and a
smaller (0.2\,--\,0.3\,mag) post-outburst brightness
decline compared to the optical.

The FUor becomes redder when fading
(Sect.~\ref{section_decline}).

The post-outburst light curves are superposed by an
oscillating pattern with a period of about 17 days
(Sect.~\ref{section_oscillations}).

For the following discussion we assume that the variations and
features in the light curves are due
to the FUor, and that the two nearby (11 and 100 AU)
companion stars found by \cite{Caratti15} play a negligible role.

\subsection{Nature of the brightness decline}
\label{section_decline}

Figure \ref{fig_cmds} shows the color-magnitude diagrams
at optical and NIR wavelengths.
Our NIR data obtained at USB after 19 December 2014  are
supplemented by the eight data points observed with
UKIRT between 2 December and 19 December 2014 by \cite{Varricatt15}.
The $J-H$ color offset between our data and the Varricatt et al. data
is not clear; it could be due to the presence of a reflection nebula
or circumstellar dust, measured with different spatial resolution.

The FUor becomes redder when fading, and
the orientation of the $A_{V}$ vector agrees with the
elongated distribution of the data points.
This also holds for other filter combinations.
The NIR data indicate $A_{V} \approx 1$.
Thus, the radiation becomes optically thick at wavelengths shorter than $V$. In addition
a reflection nebula might be present.
Both effects could explain why at optical wavelengths the
$A_{V}$ values are smaller than in the NIR (\citealt{Kruegel09}).

We suggest that
the brightness decline between October 2014 and April 2015 is
due to increasing extinction caused by dust swirled up by the outburst.
Then after April 2015 the extinction growth stopped;
one may speculate that during the further evolution of
the system holes will be blown into the dust clouds
leading to a subsequent re-brightening.

\subsection{Nature of the brightness oscillations}
\label{section_oscillations}

The post-outburst light curves are superposed by an
oscillating pattern, most visible in Fig.~\ref{fig_all_lc}
in the $r$ and $i$ bands
during the first 50 days. When zooming in,
the oscillations are recognizable throughout the entire
$r$, $i$, and $z$ light curves.

\begin{figure}
  \vspace{-4mm}
  \includegraphics[width=\columnwidth, angle=0]{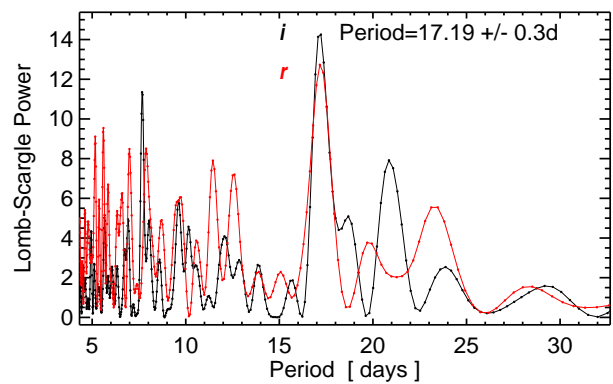}
  \caption{Lomb-Scargle periodogram of the light curve oscillations
    in the $i$ and $r$ band
    after removal of the long-term trend via a third order polynomial.
    The peak at 17.19 days (determined from a Gaussian fit)
    has a width of $\sigma$\,=\,0.3 days, and
    a false alarm probability of $\sim$0.4\%.
  }
  \label{fig_period}
\end{figure}

Applying the Lomb-Scargle method (\citealt{Lomb76}, \citealt{Scargle82}) to the entire
$r$, $i$, and $z$ light curves reveals a
local peak at about 17 days,
alongside strong power from the long-term decline
(in Fig.~\ref{fig_all_lc}).
To get rid of this power,
we removed the long-term trend
trying several fits:
a sine function,
a third-order polynomial,
and a piece-wise linear function consisting of three parts:
the high state, the decline, and the transition plateau.
For all types of long trend removal,
we find a period of about 17.2 days.
This period agrees with the results using phase dispersion minimization (PDM, \citealt{Stellingwerf78}) and the
technique to find minima in the phase-folded light curves
by \cite{Lafler65}.
We note that both Lomb-Scargle and Lafler-Kinman yield
exactly the same resulting period (Fig.~\ref{fig_period}).

\begin{figure*}
  \vspace{-2mm}
  \includegraphics[width=0.67\columnwidth, angle=0]{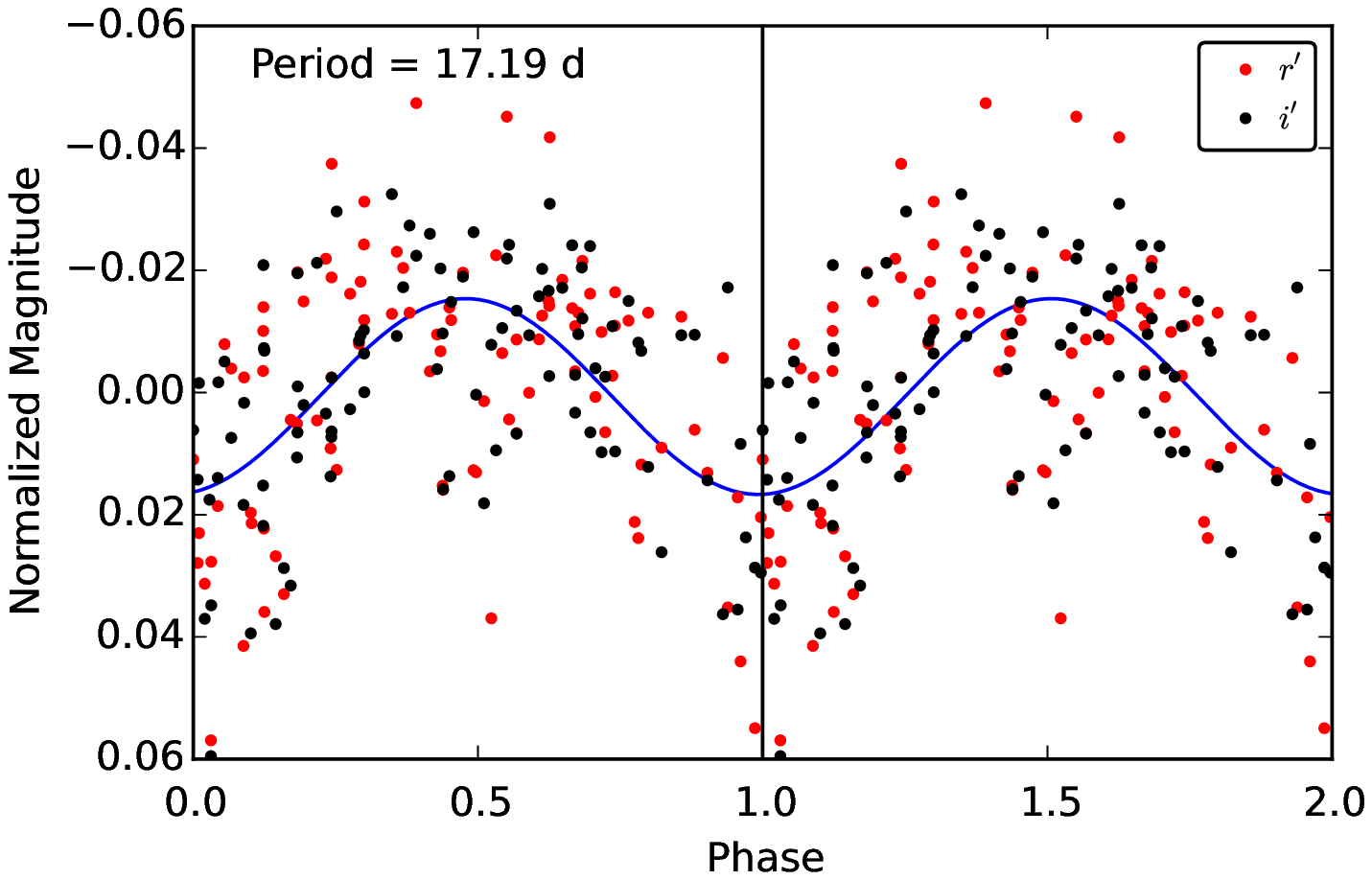}
  \includegraphics[width=0.67\columnwidth, angle=0]{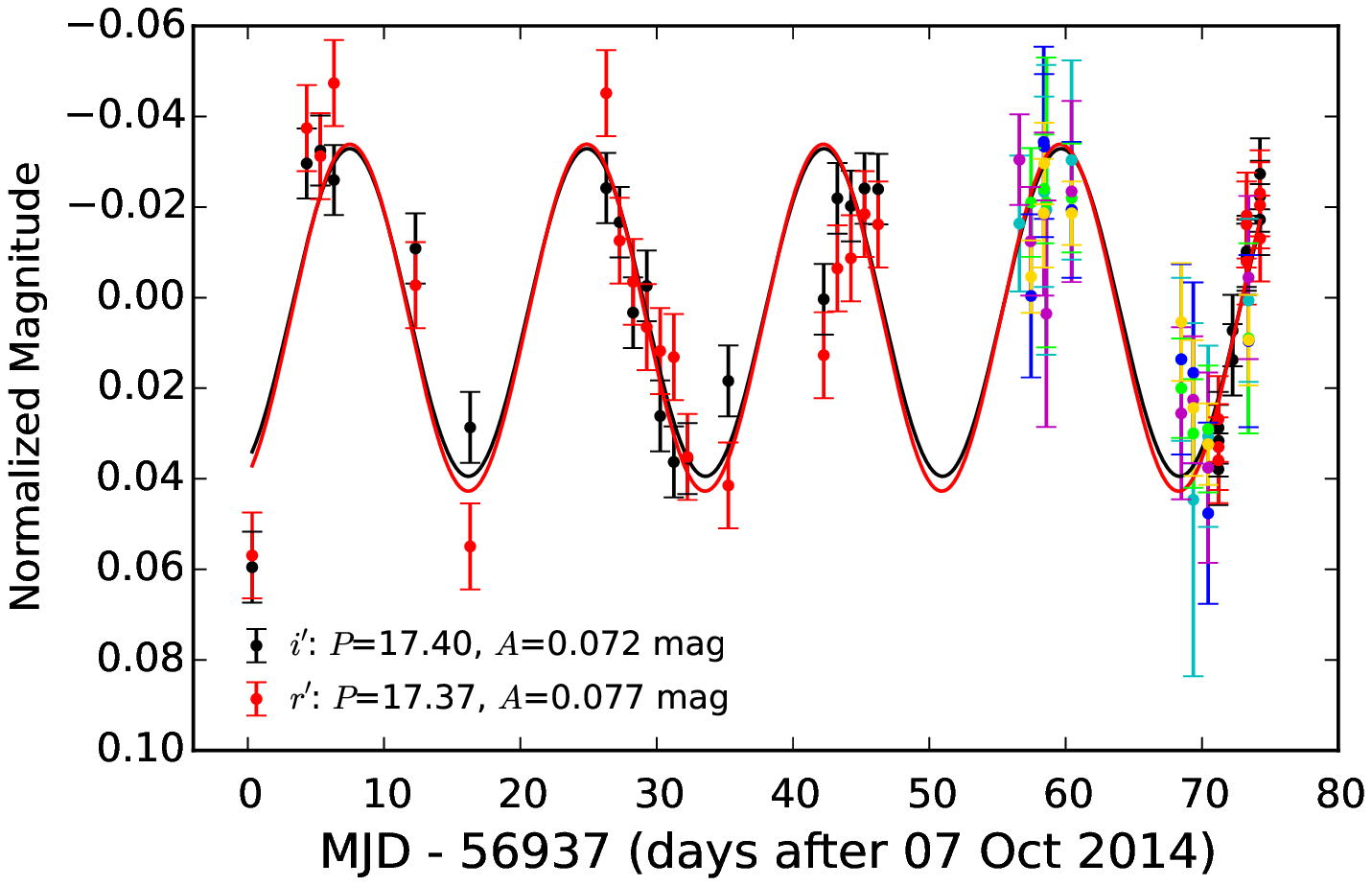}
  \includegraphics[width=0.67\columnwidth, angle=0]{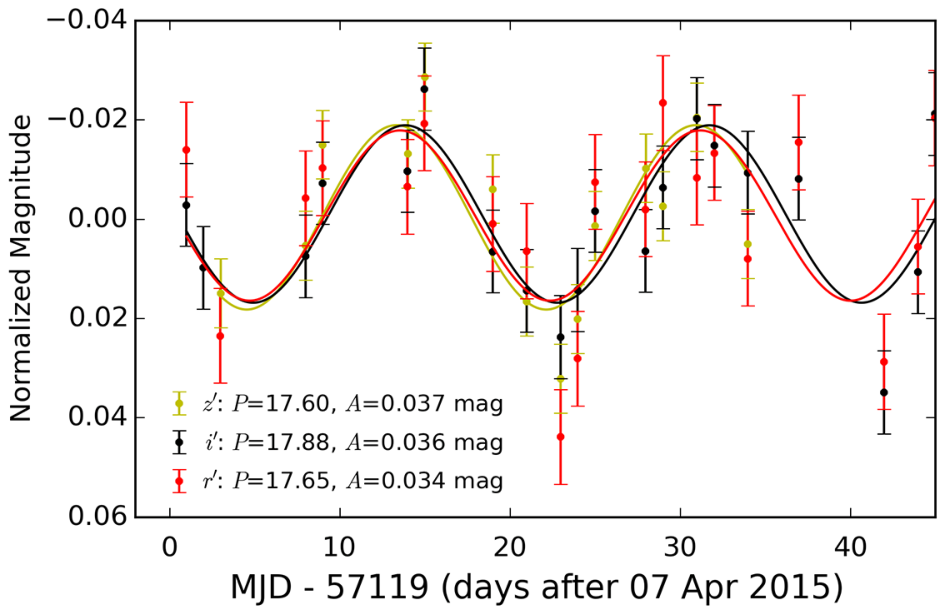}
  \caption{
    {\it Left:} Phase curves in $r$ and $i$
    derived with the Lafler-Kinman algorithm
    after removal of the long-term trend
    via a piece-wise linear function.
    Overplotted is a least squares fitted sine function (blue);
    the $\chi^2$ values are $\sim$3.6 for the sine function
    and $\sim$5.2 for a constant function at zero mag.
    {\it Middle and right:}
    Light curves in the high state
    and the transition state
    after removal of the long-term trend
    via a piece-wise linear function.
    Overplotted on each light curve
    are fitted sine functions of period and amplitude as labeled.
    The $\chi^2$ values are $\sim$2.2 for the sine functions, and
    for a constant function at zero mag $\sim$10 {\it (middle)}
    and $\sim$4 {\it (right)}.
    The middle
    panel also shows the UKIRT data of \cite{Varricatt15};
    $Z,Y,J,H,$ and $K$ in magenta, cyan, green, blue, and yellow, respectively.
  }
  \label{fig_oscillation}
\end{figure*}

We did not find any significant change of the period with time.
Likewise, the amplitude of the oscillations appears wavelength
independent for $r,i,z$, the three bands where it could be measured
reliably.
To check whether this is also supported by the longer wavelength NIR data,
the UKIRT data taken during eight nights in December 2014
by \cite{Varricatt15} turn out to provide valuable clues.
They are overplotted in Fig.~\ref{fig_oscillation} (top).
Obviously, these NIR data fit the same model as the optical data do.
Secondly, the UKIRT data points
in the  $JHK$ color-color diagram show no elongation in the direction of the $A_{V}$ vector (Fig.~\ref{fig_cmds}).
These two findings argue in favor of a wavelength
independence of the amplitude of the oscillations.

The amplitude of the oscillations declines
from $\sim$0.08\,mag in October 2014 to $\sim$0.04\,mag in May
2015 (Fig.~\ref{fig_oscillation}).

In search for explanations for the brightness oscillations we
considered the following scenarios:

  Pulsations of the accreting star: $\delta$ Cep and RR Lyr stars show
  strongly asymmetric saw-tooth light curve profiles, in contrast to V960\,Mon,
  which shows rather symmetric profiles.

  Eclipsing events: \cite{Pooley15} report on low X-ray absorption and
  \cite{Caratti15} inferred both a low
  extinction towards the FUor and a disk inclination of about
  23$^{\circ}$, i.e., more face-on than edge-on.
  This does not support eclipsing events, neither by a companion nor
  by dust clouds.
  In addition, the color-neutrality of the oscillations argues
  against eclipses by rotating dust clouds, such as proposed for V1647 Ori
  by \cite{Acosta-Pulido07}.

  Rotating hot spots fed by magnetic tubes: In this case a bluer when brighter
  behavior is expected, in contrast to
  the color-neutrality of the oscillations (e.g., \citealt{Herbst94}).

  Flickering: The light curves should show a redder when brighter
  behavior as observed in FU Orionis itself (\citealt{Kenyon00, Siwak13}),
  in contrast to our observations, hence questioning this explanation.

  An orbiting accreting hot Jupiter (\citealt{Clarke03}):
  If the planet and disk
  planes are not perfectly aligned, there are geometries for which
  the ``hot spot'' signature would be more detectable when the planet is
  approaching the observer than on the opposite phase (\citealt{Powell12}).
  However, such a scenario predicts a bluer when brighter behavior,
  in contrast to our observations.

  Finally, the faint companion at 11 AU found by \cite{Caratti15} is too far off to exhibit an orbit of $\sim$17 days.

Obviously, none of these scenarios provides a really satisfactory
explanation for the brightness oscillations.
Therefore, we now consider a
different picture of a close binary surrounded by a circumbinary disk,
as discussed and simulated by \cite{Artymowicz96} and refined by,
e.g., \cite{Guenther02} and \cite{DeVal-Borro11}.
These simulations show that the rotating binary creates a gap in the
circumbinary disk.
The gap is not empty and matter flows in a confined stream
from the circumbinary disk towards each
star, feeding its small circumstellar accretion disk.
Depending on the parameters of the system,
the binary orbit can become eccentric.
Then the interaction of the two circumstellar accretion disks and
subsequent accretion rate onto the two stars
may vary periodically, being stronger at periastron than at
apoastron.
This kind of model was successfully applied to several
regularly accreting T\,Tauri binaries with circumbinary disks, such as DG Tau and GG Tau (\citealt{Guenther02, DeVal-Borro11}).

To conclude, a close binary with eccentric orbit could naturally explain the observed brightness oscillations of the present FUor -- a scenario we therefore strongly suggest.

\newpage

\begin{acknowledgements}

  This work is supported by the
  Nordrhein-Westf\"alische Akademie der Wissenschaften und der K\"unste
  in the framework of the academy program of the Federal Republic of
  Germany and the state Nordrhein-Westfalen, by
  Deutsche Forschungsgemeinschaft (DFG HA3555/12-1)
  and by Deutsches Zentrum f\"ur Luft-und Raumfahrt (DLR 50\,OR\,1106).
  This work was partly supported by the Hungarian Research Fund OTKA K101393.
  This work was supported by the Momentum grant of the MTA CSFK
  Lend\"ulet Disk Research Group.
  Construction of the IRIS infrared camera was supported by the National
  Science Foundation under grant AST07-04954.
  We thank Lynne Hillenbrand for her constructive and helpful referee report.

\end{acknowledgements}


\bibliographystyle{aa}
\bibliography{FUOR_J0659_AA_V10}

\end{document}